\documentclass[10pt, superscriptaddress,twocolumn,showpacs,prb,aps]{revtex4-1}

\usepackage{graphicx}
\usepackage{dcolumn}
\usepackage{bm}
\usepackage{xspace}
\usepackage{multirow}
\usepackage{natbib}
\usepackage{color}
\usepackage{makecell}

\newcommand{\RUC}{Department of Physics, Renmin University of China, Beijing 100872, China}
\newcommand{\Columbia}{Department of Applied Physics and Applied Mathematics, Columbia University, New York 10027, USA}
\newcommand{\BESSY}{Institute for Solid State Research, IFW Dresden, Dresden 01171, Germany}
\newcommand{\Rutgers}{Rutgers Center for Emergent Materials and Department of Physics and Astronomy, Rutgers University, New Jersey 08854, USA}
\newcommand{\IOP}{Beijing National Laboratory for Condensed Matter Physics, and Institute of Physics, Chinese Academy of Sciences, Beijing 100190, China}
\newcommand{\CenterQM}{Collaborative Innovation Center of Quantum Matter, Beijing, China}
\newcommand{\BJLab}{Beijing Key Laboratory of Opto-electronic Functional Materials $\&$ Micro-nano Devices (Renmin University of China)}
\newcommand{\Columbiax}{Department of Electrical Engineering, Columbia University, New York 10027, USA}
\newcommand{\Ukraine}{Taras Shevchenko National University of Kyiv, Kyiv 01601, Ukraine}
\newcommand{\Shanghai}{State Key Laboratory of Functional Materials for Informatic, SIMIT, Chinese Academy of Sciences, Shanghai 200050, China}

\newcommand{\BIS}{Bi$_{2-x}$In$_{x}$Se$_{3}$}
\newcommand{\EF}{$E_F$}

\begin{document}

\title{Sudden gap-closure across the topological phase transition in {\BIS}}

\author{Rui Lou}
\thanks{These two authors contributed equally to this work.}
\affiliation{\RUC}

\author{Zhonghao Liu}
\thanks{These two authors contributed equally to this work.}
\affiliation{\BESSY}
\affiliation{\Shanghai}

\author{Wencan Jin}
\affiliation{\Columbia}

\author{Haifeng Wang}
\author{Zhiqing Han}
\author{Kai Liu}
\affiliation{\RUC}

\author{Xueyun Wang}
\affiliation{\Rutgers}

\author{Tian Qian}
\affiliation{\IOP}

\author{Yevhen Kushnirenko}
\affiliation{\BESSY}
\affiliation{\Ukraine}

\author{Sang-Wook Cheong}
\affiliation{\Rutgers}

\author{Richard M. Osgood, Jr.}
\affiliation{\Columbia}
\affiliation{\Columbiax}

\author{Hong Ding}
\affiliation{\IOP}
\affiliation{\CenterQM}

\author{Shancai Wang}
\email{scw@ruc.edu.cn}
\affiliation{\RUC}
\affiliation{\BJLab}

\begin{abstract}
  The phase transition from a topological insulator to a trivial band insulator
  is studied by angle-resoled photoemission spectroscopy on {\BIS} single crystals.
  We first report the complete evolution of the bulk band structures throughout the
  transition. The robust surface state and the bulk gap size ($\sim$ 0.50 eV) show
  no significant change upon doping for $x$ = 0.05, 0.10 and 0.175. At $x$ $\geq$ 0.225,
  the surface state completely disappears and the bulk gap size increases, suggesting
  a sudden gap-closure and topological phase transition around $x \sim$ 0.175$-$0.225.
  We discuss the underlying mechanism of the phase transition, proposing that it is
  governed by the combined effect of spin-orbit coupling and interactions upon band
  hybridization. Our study provides a new venue to investigate the mechanism of the
  topological phase transition induced by non-magnetic impurities.
\end{abstract}


{\maketitle}

Topological insulators (TIs), principally three-dimensional (3D) topological crystals, have attracted
much attention and led to an upsurge in finding new topological phases of matter. Extensive theoretical
and experimental work has been carried out on both the robust surface state (SS) and topological properties
of the bulk bands. This body of work now allows distinguishing a TI from a ``trivial" band insulator.\cite{
Hasan2010,Qi2011,Fu2007a,Moore2007,Hsieh2009,Chen2009,Hsieh2009a,Xia2009,Hsieh2008,Roy2009,Roushan2009,
Dziawa2012,Wojek2014,Pan2011,Nechaev2013,Scholz2012} Further, the mechanism of the transition between these
two classes of insulators, found by substituting non-magnetic impurities, has been proposed to be a homogeneous
3D topological phase transition (TPT) scenario, $i.e.$, the so-called linear gap-closure scenario, suggesting
that the bulk gap is determined by the spin-orbit coupling (SOC) alone, and thus would decrease monotonically
along with the decreasing SOC strength in the non-trivial phase. This transition happens when the bulk gap
closes and an inversion of the bulk conduction band (CB) and valance band (VB) occurs.\cite{Hasan2010,Qi2011,
Zhang2009,Mosfets2006,Fu2007,Liu2010}

Practically, the 3D TPT induced by non-magnetic impurities could only be realized in few real
systems, mainly TlBi(S$_{1-x}$Se$_{x}$)$_{2}$ and {\BIS}. For TlBi(S$_{1-x}$Se$_{x}$)$_{2}$,
the existence of a critical point between the TI TlBiSe$_{2}$ and the trivial metal TlBiS$_{2}$
is observed.\cite{Sato2011,Xu2011,Chen2010,Souma2012,Xu2015} Upon doping, an unexpected surface
bandgap, $i.e.$, the Dirac gap, is observed near the TPT.\cite{Sato2011,Souma2012,Xu2015} For
{\BIS}, a broader TPT from TI to trivial insulator is reported, as suggested by transport and
photoemission measurements on {\BIS} thin films.\cite{Brahlek2012,Wu2013,Watanabe1989} This
linear gap-closure scenario motivates a study to characterize the evolution of the bulk bands
of these two systems so as to fully understand the TPT.\cite{Zhang2009,Mosfets2006} While the
SS has been characterized, observation of these bulk band features have not yet been reported.
Therefore, to date, the underlying mechanism of non-magnetic-impurity-induced TPT still remains
elusive. In order to obtain a much clearer insight into the mechanism, we performed systematic
high-resolution angle-resoled photoemission spectroscopy (ARPES) measurements on {\BIS} single
crystals, focusing on the detailed evolution of both the bulk bands and the SS during the TPT.

In this paper, we report an observation of the phase transition from a TI to a trivial
insulator in {\BIS} single crystals with various nominal doping levels ($x$ = 0.05, 0.10,
0.175, 0.225 and 0.30). We demonstrate that the bulk gap size ($\sim$ 0.50 eV) shows no
significant change in the topologically non-trivial region ($i.e.$, $x$ = 0.05, 0.10 and
0.175), instead of a linear gap-closure behavior. The bulk gap appears to abruptly close
at a specific doping level ($\sim$ 0.175$-$0.225) accompanied by the complete suppression
of the SS. After the TPT, the bulk gap size increases for $x$ = 0.225 and 0.30. Both the
SOC and interactions upon band hybridization are suggested to cooperate in this local
phase transition.

High quality single crystals of {\BIS} were grown by slowly cooling a stoichiometric
mixture of high purity elements of Bismuth, Indium and Selenium in an evacuated quartz
tube.\cite{Analytis2010} ARPES measurements were performed at Renmin University of
China and Institute of Physics, Chinese Academy of Sciences, with a He-discharge lamp
and at the 1-cubed ARPES end station at BESSY using synchrotron radiation. The overall
angular and energy resolutions are better than 0.2{$^\circ$} and 5 meV, respectively.
Samples were cleaved {\emph{in situ}} yielding flat (001) surfaces, and measured at
$T$ $\sim$ 10 K, with a pressure better than 4 $\times$ 10$^{-11}$ Torr.

\begin{figure}[htb]
  \begin{center}
    \includegraphics[width=0.85\columnwidth]{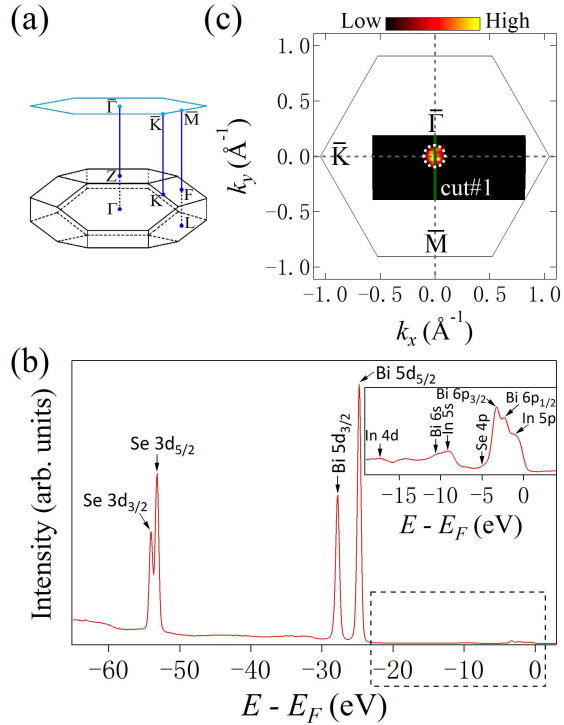}
  \end{center}
  \vspace{-1.55em}
  \caption{(Color online)
  (a) Bulk BZ and its surface projection of {\BIS}.
  (b) Core level photoemission spectrum ($h\nu$ = 100 eV) of $x$ = 0.05. The inset is zoom
      in of the VB from the dashed box.
  (c) ARPES intensity plot ($h\nu$ = 70 eV) of $x$ = 0.05 at {\EF} as a function of the 2D
      wave vector. The intensity is obtained by integrating the spectra within $\pm$15 meV
      with respect to {\EF}. Cut\#1 indicates the $\bar{\Gamma}\bar{\mathrm{M}}$ direction,
      along which the data are presented in Fig.~\ref{fig2}. The white dashed circle is guide
      to the eyes, served as the FS.
  }\label{fig1}
\end{figure}

The schematic bulk Brillouin zone (BZ) of {\BIS} is presented in Fig.~\ref{fig1}(a). As is shown
in Fig.~\ref{fig1}(b), well-defined peaks in the core level photoemission spectrum of $x$ = 0.05
demonstrate the high quality of the series of crystals used in this work. One can obtain insight
of the binding energy within $\sim$ 20 eV below {\EF} in the core level spectrum, as shown in the
inset of Fig.~\ref{fig1}(b). Fig.~\ref{fig1}(c) shows the Fermi surface (FS) mapping data of $x$
= 0.05 as a function of in-plane wave vector.

\begin{figure*}[htb]
  \begin{center}
    \includegraphics[width=1.83\columnwidth]{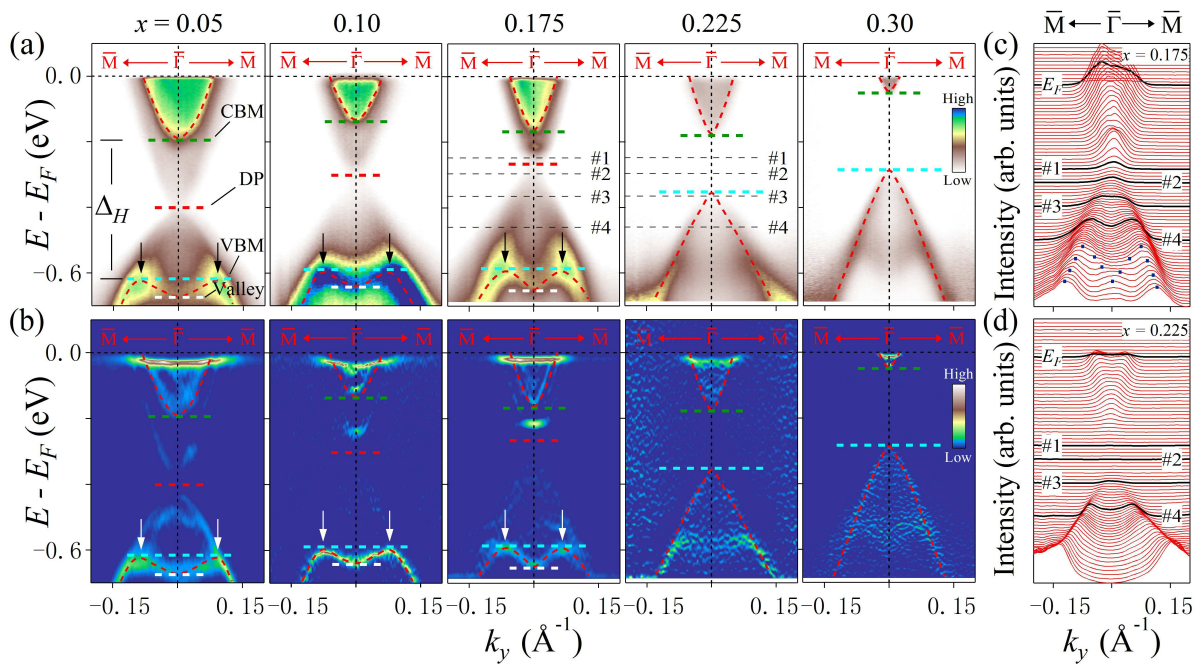}
  \end{center}
  \vspace{-1.8em}
  \caption{(Color online)
  ARPES intensity plots (a) and the corresponding second derivative plots (b) of {\BIS} along
  $\bar{\Gamma}\bar{\mathrm{M}}$ direction ($h\nu$ = 20 eV). The nominal composition value is
  noted above each representative plot. The CBM, VBM, DP, valley and $\delta$$k$'s are marked
  by green, blue, red, white dashed lines and solid arrows, respectively, the red dashed curves
  represent the bulk bands. Band dispersions are quantitatively determined by energy distribution
  curves (EDCs) and momentum distribution curves (MDCs) analysis and overlaid with the second
  derivative plots. (c) and (d) show the corresponding MDC plots of $x$ = 0.175 and 0.225 from
  (a), respectively. Spectra at the binding energies of -0.248, -0.297, -0.366 and -0.460 eV,
  as indicated by black dashes in (a), are highlighted by thick black curves. Blue dots in (c)
  are guides to the eyes for the valley structure.
  }\label{fig2}
\end{figure*}

Fig.~\ref{fig2}(a) shows the band dispersions of series of doped samples along the $\bar{\Gamma}\bar{\mathrm{M}}$
direction, indicated via cut\#1 in Fig.~\ref{fig1}(c). The corresponding second derivative plots are shown in
Fig.~\ref{fig2}(b). The data were collected using an incident photon energy of 20 eV, ensuring the $k_{z}$
positions close to the bulk $\Gamma$ point.\cite{Xia2009,Bianchi2010} Also, we verify the $k_{z}$ dispersion of
the band features by performing photon-energy-dependent measurement (see Supplemental Material Part 1).\cite{
Supplemental} The plots in Fig.~\ref{fig2}(a) clearly show the SS for $x$ $\leq$ 0.175, thus indicating a
topologically non-trivial region, and the absence of the SS for $x$ $\geq$ 0.225, thus indicating a topologically
trivial region. This could be further confirmed by the MDCs of $x$ = 0.175 and 0.225 shown in Figs.~\ref{fig2}(c)
and~\ref{fig2}(d), respectively. The purely flat curves \#1, \#2 in Fig.~\ref{fig2}(d) are in great contrast to
the well-dispersed spectra \#1--\#4 in Fig.~\ref{fig2}(c) from the linear dispersion of the SS. Due to the absence
of the SS, the spectra \#3, \#4 in $x$ = 0.225 [Fig.~\ref{fig2}(d)] could be reasonably assigned as the band
dispersions of VB, with \#3 near the valance band maximum (VBM) and \#4 deep in the VB. These band assignments are
discussed in more detail in Supplemental Material Part 1.\cite{Supplemental}

Furthermore, one can see the evolution of the bulk bands from Figs.~\ref{fig2}(a) and~\ref{fig2}(b). The
bulk gap is defined as the difference between the conduction band minimum (CBM) and the valley of VB at
$\bar{\Gamma}$ point for the topologically non-trivial region. The difference between the CBM and VBM is
used for the topologically trivial region. In the topologically non-trivial region, the magnitude of bulk
gaps reveal binding energies of 0.48, 0.50 and 0.49 eV for $x$ = 0.05, 0.10 and 0.175, respectively,
indicating no significant change upon doping. The valley structure in VB, caused by the band inversion,
gradually weakens along with the increasing doping, demonstrating the decrease of SOC strength. This is
quantitatively proven by the momentum of the bending band ($\delta$$k$) around the VBM, as marked by solid
arrows in Figs.~\ref{fig2}(a) and~\ref{fig2}(b). The absolute values of $\delta$$k$'s are 0.091, 0.073 and
0.060 \AA$^{-1}$ for $x$ = 0.05, 0.10 and 0.175, respectively. In the topologically trivial region, the
band inversion disappears and the valley structure vanishes. The increase of the direct band gap indicates
the further decrease of SOC strength.\cite{Zhang2009,Mosfets2006} Considering the evolution of both the SS
and bulk band structures, we anticipate that a local phase transition characterized by a sudden gap-closure
happens close to $x \sim$ 0.175$-$0.225. Another observation is the Dirac point (DP) moving toward the CBM
with the increasing doping. These can be expected to merge at the critical point, which also confirms the
existence of a TPT.\cite{Hasan2010,Qi2011}

\begin{figure}[htb]
  \begin{center}
    \includegraphics[width=1\columnwidth]{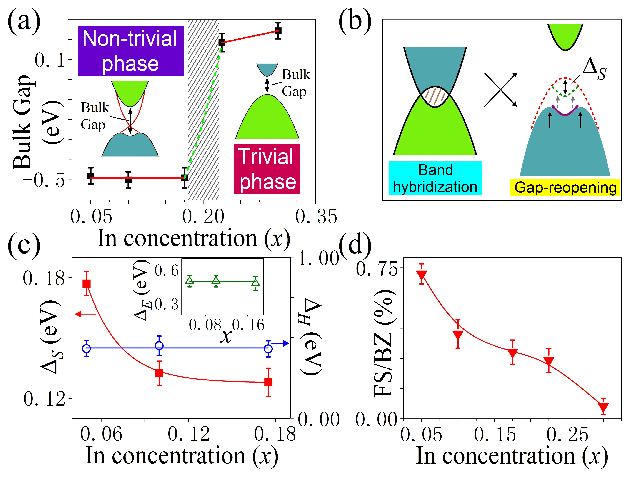}
  \end{center}
  \vspace{-1.6em}
  \caption{(Color online)
  (a) The evolution of the bulk gap size at $\bar{\Gamma}$ point. The definition of bulk gap
      in two phases are shown in the insets. The critical point is within the shadow region.
  (b) A schematic picture of bulk bands in band hybridization (left panel) and gap-reopening
      (right panel) in the topologically non-trivial phase. The $\delta$$k$'s (black vertical
      arrows), unperturbed VB (red dashed curve), unperturbed CB (purple solid curve) and
      shifted CB (green dashed curve) are indicated to enable an estimate of $\Delta_{S}$.
  (c) $\Delta_{S}$ (red solid square) and $\Delta_{H}$ (blue open circle) as a function of
      $x$ in the topologically non-trivial phase. The evolution of the estimated bulk gap,
      $\Delta_{E}$ $\approx$ $\lambda\cdot\Delta_{S}$ + $\Delta_{H}$ ($\lambda$ = 0.4) (green
      open triangle), upon doping is shown in the inset.
  (d) The doping dependence of the ratio of the enclosed FS area to the whole BZ.
  }
  \label{fig3}
\end{figure}

The extracted evolution of the bulk gap is shown in Fig.~\ref{fig3}(a). The gap size has
a dramatic transformation between $x$ = 0.175 and 0.225, as shown in the shadow region of
Fig.~\ref{fig3}(a), suggesting a critical transition. This behavior is in contrast to the
mild change for $x$ = 0.05, 0.10 and 0.175 (defined as negative values to distinguish from
that in the topologically trivial phase). After crossing this apparent critical point, a
strong increase from $x$ = 0.225 to 0.30 is observed. This bulk gap evolution deviates from
that of the linear gap-closure scenario. The evident deviation shows that the underlying
microscopic mechanism of TPT could not be simply elucidated by a SOC dominant effect. This
first systematic observation of the bulk band structure evolution suggests a very new
mechanism is present.

As proposed in Ref. 17, a band inversion is induced by SOC for a 3D TI. In the topologically non-trivial
region, after turning on the SOC, there is a downward and upward shift of the CB and VB, respectively,
causing the band inversion and they hybridize with each other around $\bar{\Gamma}$ point. Further, the
bulk gap would be expected to reopen owing to the interactions within the overlap. Meanwhile, our observed
bulk gap behavior also indicates that the reopened bulk gap at $\bar{\Gamma}$ point ($i.e.$, the gap between
the CBM and the valley) is no longer dominated by SOC alone. Therefore, the combined effect of both SOC and
interactions upon band hybridization should be taken as the determinant during the TPT.

It is possible to estimate the relative contribution from SOC and band hybridization. The overlap ratio
($\Delta_{S}$) between the CB and VB after turning on the SOC is, $\Delta_{S}$ = $\vert E_{g} - E_{SOC}
\vert$, where E$_{g}$ is the energy gap between the CBM and VBM before turning on the SOC and proportional
to the covalent bonding strength within the quintuple layer,\cite{Slater1954} and E$_{SOC}$ is the total
energy shift of the CB and VB caused by the SOC. As the $\delta k$'s of VB are exactly the momentum of
the crossing points between the CB and VB when overlapping, thus, after the reopening, the indirect gap
between the CBM and VBM [as shown in Fig.~\ref{fig2}(a)] is defined as the hybridization gap ($\Delta_{H}$).
Therefore, we suggest that, $\Delta_{S}$, which corresponds to the relative strength of SOC, and $\Delta_{H}$,
which characterizes the interactions within the overlap, together modulate the evolution of bulk gap in the
topologically non-trivial region. In the topologically trivial region, the band inversion vanishes and along
with the band hybridization ($\Delta_{H} \equiv 0$). As a consequence, the increase of the bulk gap size is
directly attributed to the decrease of SOC strength, $i.e.$, $\Delta_{S}$, upon In doping.\cite{Zhang2009,
Mosfets2006}

As is schematically illustrated in Fig.~\ref{fig3}(b), we estimate $\Delta_{S}$ as follows: first,
extracting the $\delta k$'s of VB, shown as black vertical arrows; second, extrapolating the high
energy band dispersion to recover the unperturbed VB, and fitting the valley structure to get the
unperturbed CB, shown as red dashed and purple solid curves, respectively; last, by shifting the
unperturbed CB upward and matching the crossing points, $\delta$$k$'s, between the unperturbed CB
and VB, we recover the overlap as the shadow area in the left panel. The energy difference between
the extreme values of the red and green dashed curves is defined as $\Delta_{S}$. (See more details
in Supplemental Material Part 2.\cite{Supplemental})

In Fig.~\ref{fig3}(c), the evolution of $\Delta_{S}$ and $\Delta_{H}$ are plotted as a function of
In concentration. They have remarkably different tendencies in the topologically non-trivial phase,
in which, upon In doping, $\Delta_{S}$ decreases monotonically, while $\Delta_{H}$ shows no noticeable
change. We now discuss the possible combination of $\Delta_{S}$ and $\Delta_{H}$ in determining the
bulk gap size and thus try to understand the underlying mechanism of TPT. From a simple perspective,
the bulk gap value is the summation of $\Delta_{H}$ and the difference between the VBM and the valley.
Therefore, a reasonable estimate of the bulk gap ($\Delta_{E}$) using the experimental $\Delta_{S}$
and $\Delta_{H}$ values is, $\Delta_{E}$ $\approx$ $\lambda\cdot\Delta_{S}$ + $\Delta_{H}$, in which
the factor $\lambda$ characterizes the contribution from SOC. In the topologically non-trivial region,
$\lambda$ = 0.4, and in the topologically trivial region, $\lambda$ = 1.0. By noticing that $\Delta_{H}
\sim$ 0.4 eV and $\Delta_{S} \sim$ 0.1$-$0.2 eV in the topologically non-trivial phase, the gap size
is seen to be dominated by $\Delta_{H}$ according to this formula. As a result, although $\Delta_{S}$
decreases faster along with increasing doping, the gap size changes slightly [as presented in the inset
of Fig.~\ref{fig3}(c)] at low In concentrations until the band inversion vanishes. Beyond the critical
point, $\Delta_{H}$ vanishes, thus the bulk gap is directly determined by the SOC strength, which
decreases monotonically along with the increasing In concentration. This scenario reasonably addresses
our observed bulk gap evolution, and offers an important insight into the mechanism of TPT. Considering
the behavior of $\Delta_{H}$, we give a reasonable speculation here. According to the first-principles
calculations of {\BIS} proposed in Ref. 33, in the topologically trivial region, the CBM is composed of
the Bi 6$p$ and In 5$s$ orbitals. Both of these orbitals are involved in the overlap between the CB and
VB in the topologically non-trivial region. As the interactions upon band hybridization are proportional
to the density of states (DOS) within the overlap, and considering the DOS is intimately connected with
the overlap ratio, $\Delta_{H}$ may be expected to decrease along with the decreasing $\Delta_{S}$ in
common sense. However, our observed nearly invariant $\Delta_{H}$ possibly indicates the non-negligible
contribution of the increasing In 5$s$ orbital in modulating the DOS within the overlap.

We estimate the carrier concentrations of our samples by calculating the ratios of the enclosed
FS area to the whole BZ and present them in Fig.~\ref{fig3}(d). This ratio decreases monotonically
with increasing doping, agreeing well with the transport results proposed in Ref. 26.

\begin{figure}[htb]
  \begin{center}
    \includegraphics[width=1\columnwidth]{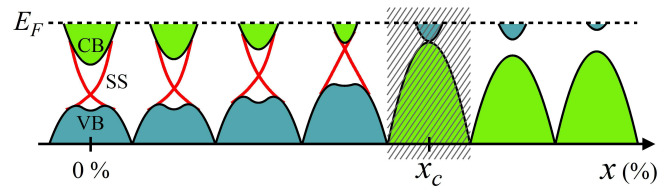}
  \end{center}
  \vspace{-1.8em}
  \caption{(Color online)
  A schematic picture of the band structure evolution of {\BIS} as a function of In concentration.
  The different colors of the bulk bands represent different orbital characters and parities.
  }
  \label{fig4}
\end{figure}

A schematic picture of the band structure evolution during the TPT is conceptually
sketched in Fig.~\ref{fig4}. In the topologically non-trivial region ($x < x_{c}$),
the band inversion exists, thus the interactions upon band hybridization and SOC
together modulate the bulk gap size. As a result, the bulk gap shows no significant
change. With the increase of In doping, the overlap ratio between the CB and VB
decreases. At the critical point ($x = x_{c}$), the band inversion vanishes and so
does the band hybridization, the bulk gap collapses accompanied by the vanishing of
the SS. In the topologically trivial region ($x > x_{c}$), the further decreasing
SOC strength separates the CB and VB, giving rise to the increase of bulk gap. The
DP moves toward the CBM with the increasing doping, and the extrapolated merging
point is at the vicinity of $x_c$. This behavior can be explained as a result of
the reduction of the inverted band overlap ratio between the CB and VB.

To summarize, we have performed ARPES experiments on {\BIS} single crystals to study the band structure evolution
of the TPT induced by non-magnetic impurities. We report the evolution of the bulk bands throughout the transition,
and propose a sudden gap-closure behavior across the phase transition, instead of the linear gap-closure scenario
dominated by SOC alone. Our study suggests that the interactions upon band hybridization and SOC together determine
the TPT, providing a novel perspective on the underlying mechanism.

We thank Rong Yu, Xi Dai for helpful discussions and Hechang Lei, Rui Cao, Zongyao Zhang for the
help in X-ray diffraction measurements. This work was supported by grants from National Science
Foundation of China, National Basic Research Program of China (973 Program), Ministry of Education
of China, China Academic of Science and SSSTC. Wencan Jin and Richard M. Osgood, Jr. were supported
by the US Department of Energy, Office of Basic Energy Sciences, Division of Materials Sciences and
Engineering under Award Contract No. DE-FG 02-04-ER-46157. The work at Rutgers was funded by the
Gordon and Betty Moore Foundation's EPiQS Initiative through Grant GBMF4413 to the Rutgers Center
for Emergent Materials.

\end{document}